\documentclass[aps,prd,epsf,twocolumn,showpacs,groupedaddress]{revtex4}
\usepackage{graphicx}
\usepackage{amsmath,amssymb,latexsym}
\usepackage{bm}
\newcommand{\bwide}{\begin{widetext}}
\newcommand{\ewide}{\end{widetext}}
\newcommand{\beq}[1]{\begin{equation} \label{(#1)}}
\newcommand{\eeq}{\end{equation}}
\newcommand{\ba}[1]{\begin{eqnarray} \label{(#1)}}
\newcommand{\ea}{\end{eqnarray}}

\def\Journal#1#2#3#4{{#1} {\bf #2}, #3 (#4)}

\def\beq{\begin{equation}}
\def\eeq{\end{equation}}

\def\NPB{{\em Nucl. Phys.} B}
\def\PLB{{\em Phys. Lett.}  B}
\def\PRL{\em Phys. Rev. Lett.}
\def\PRD{{\em Phys. Rev.} D}

\begin{document}
\title{Probing mSUGRA via the Extreme Universe Space Observatory}
\author{Luis Anchordoqui}\email[]{l.anchordoqui@neu.edu}
\affiliation{Department of Physics, Northeastern University, Boston MA 02155}
\author{Haim Goldberg}\email[]{goldberg@neu.edu}
\affiliation{Department of Physics, Northeastern University, Boston MA 02155}
\author{Pran Nath}\email[]{nath@neu.edu}
\affiliation{Department of Physics, Northeastern University, Boston MA 02155}

\begin{abstract}
An analysis is carried out within mSUGRA of the estimated number
of events originating from upward moving ultra-high energy
neutralinos passing through Earth's crust that could be detected
by the Extreme Universe Space Observatory (EUSO).  The analysis
exploits a recently proposed technique that differentiates
ultra-high energy neutralinos from ultra-high energy neutrinos
using their different absorption lengths in the Earth's crust. It
is shown that for the  part of the parameter space, where the
neutralino is mostly a Bino and with squark mass $\sim 1$ TeV,
EUSO could see ultra-high energy neutralino events within mSUGRA
models with essentially no background. In the energy range
 $10^{9}~{\rm GeV} < E_{\tilde \chi}
< 10^{11}~{\rm GeV}$ the unprecedented aperture of EUSO makes the
telescope sensitive, after 3~yr of observation, to neutralino
fluxes as low as  $d\Phi/dE_{\tilde \chi} > 1.1 \times 10^{-6}\
(E_{\tilde \chi}/{\rm GeV})^{-1.3}$~GeV$^{-1}$ cm$^{-2}$\
yr$^{-1}$\ sr$^{-1},$ at the 95\% CL. Such a hard spectrum is
characteristic of supermassive particles' $N$-body hadronic
decay. The case in which the flux of ultra-high energy
neutralinos is produced via decay of metastable heavy ($m_X = 2
\times 10^{12}$~GeV) particles with uniform distribution
throughout the universe, and primary decay mode into 5 quarks + 5
squarks, is analyzed in detail. The normalization of the ratio of
the relics' density to their lifetime has been fixed so that the
baryon flux produced in the supermassive particle decays
contributes to about 1/3 of the events reported by the AGASA
Collaboration below $10^{11}$~GeV, and hence the associated GeV
$\gamma$-ray flux is in complete agreement with EGRET data. For
this particular case, EUSO will collect between 4 and 5
neutralino events (with 0.3 of background) in $\approx 3$~yr of
running. NASA's planned  mission, the Orbiting Wide-angle
Light-collectors (OWL), is also briefly discussed in this context.
\end{abstract}
\pacs{11.30.Pb, 96.40.Pq -- NUB-3247/Th-04}
\maketitle

\section{Introduction}

mSUGRA~\cite{msugra}  and its extensions
(generically called SUGRA models) are currently
the leading candidates for physics beyond the standard model. These models contain a
consistent mechanism for the breaking of supersymmetry softly by gravity
mediation.
An attractive feature of these models is that with
R parity conservation the lightest neutralino is a possible candidate
for cold dark matter~\cite{goldberg} in a significant part of the
mSUGRA parameter space~\cite{scaling}.
 Further, over most of the parameter space
the phenomenon of scaling occurs~\cite{scaling} so that the light
neutralino is mostly
the supersymmetric partner of the $U(1)_Y$ gauge boson $B_{\mu}$, i.e.,
it is mostly a $U(1)_Y$ gaugino or a Bino~\cite{scaling,roberts}.  The
parameter space of mSUGRA is
characterized by the universal scalar mass, $m_0$, the
universal gaugino mass, $m_{\frac{1}{2}}$, the universal
trilinear coupling, $A_0$
(all taken at the grand unification scale $M_G\sim 2\times 10^{16}$ GeV),
$\tan\beta =<H_2>/<H_1>$ where $H_2$ gives mass to the up quark, and
$H_1$ gives mass to the down quark and the lepton. In addition
the model contains the Higgs mixing parameter $\mu$ which enters in
the superpotential in the form $\mu H_1H_2$. The magnitude of $\mu$
is determined by the constraint of radiative electro-weak symmetry
breaking in the theory while $sign\mu$ is arbitrary and must be
constrained by experiment.

mSUGRA has been put to stringent test by the recent precision
data from the  satellite experiment, the Wilkinson Microwave
Anisotropy Probe (WMAP) which imposes a narrow range for cold dark
matter (CDM) so that~\cite{bennett,spergel}
 $  \Omega_{\rm CDM} h^2 =0.1126^{+0.008}_{-0.009}$. The
candidacy of neutralinos as the dark matter of the universe is
based on relic densities surviving annihilation processes of
non-relativistic particle. Detailed analyses show that mSUGRA
allows for a small amount of the parameter space in agreement
with WMAP observations~\cite{wmap1}.  As a consequence, the
applicability of mSUGRA demands that the contribution of other
sources of CDM to the dark matter mix are negligible.  In this
paper, we will be interested in a flux of ultrarelativistic
neutralinos  resulting from decays of a population of CDM
metastable superheavy
particles~\cite{Berezinsky:1997hy,Kuzmin:1997cm}. In concert with
the previous statement, these  particles should contribute
negligibly to the dark matter density.

The weak couplings of neutralinos imply an interaction length in
air which is greater than the atmospheric depth, even at
horizontal incidence. The interaction probability is then
roughly  uniform throughout the atmosphere. As with neutrinos,
showers initiated by neutralino primaries can be distinguished
from hadronic events by restricting the zenith angle space to
near horizontal -- this maximizes the probability to detect
showers of weakly interacting primaries, while screening out the
electromagnetic component of hadronic showers which are initiated
high in the atmosphere. However, deeply developing neutralino
cascades cannot be isolated from neutrino induced air showers.

In this paper we show that the part of the parameter space where
the neutralino is mostly a Bino and the mass $m_{\tilde q}$ of the
first and second generation squarks is $\sim$ 1 TeV can lead to
ultra-high energy neutralino signals that may be seen by the
Extreme Universe Space Observatory
(EUSO)~\cite{Catalano:mm,Scarsi:fy}. These two conditions are
fully compatible with the WMAP constraint, and the neutralino as
lightest supersymmetric particle. Further, with appropriate cuts
the background events arising from ultra-high energy neutrinos
are essentially negligible. We discuss now the details of the
analysis.

The problem of discriminating between neutralino and neutrino
induced showers with space-based experiments has been examined
recently~\cite{Barbot:2002et}. The method makes use of the Earth
as a filter. Specifically, in the  region of the mSUGRA parameter
space under consideration the cross section for neutralino-nucleon
interaction is smaller than that for neutrino-nucleon scattering
processes. Thus, by restricting the angular bin for arrival of
upward going showers to a region where neutrinos are largely
absorbed during traversal in the Earth, it may be possible to
obtain a clean signal~\cite{Anchordoqui:2001cg}. In
Ref.~\cite{Barbot:2002et}, the discussion was presented in terms
of neutralino-nucleon cross sections parameterized as a series of
constant fractions of the neutrino-nucleon cross section. In this
paper, we first calculate the neutralino cross section in the
squark-resonance approximation.  We then proceed to estimate the
sensitivity of EUSO  to neutralino-induced air showers. The
sensitivity will be characterized by a lower bound on the
neutralino flux, which is then related to some particular models
of $X$-particle decay.

\section{EUSO outline}

In view of a mission starting in 2006-2007, the Extreme Universe
Space Observatory~\cite{Catalano:mm,Scarsi:fy} has been designed
to observe the complex relativistic cascades induced by the
incoming extra terrestrial radiation using sensors in the
ultraviolet band (300 - 400 nm), a technique pioneered by the
Fly's Eye experiment~\cite{Baltrusaitis:mx}. The telescope will
operate for more than 3 yr on board of the International Space
Station. The eye will be equipped with wide-angle Fresnel optics
lens that provide a field-of-view of $\pm 30^\circ$, at an orbit
altitude of about 400~km. The monocular stand-alone configuration
of the instrument will cover an area of $\approx 1.6 \times
10^5$~km$^2,$ imaging an air target mass that exceeds $10^{12}$
tons. This corresponds to a water equivalent (w.e.) effective
volume of $\approx 2400$~km$^3$. Observations can only be done on
clear moonless nights. Limitations associated with the cloud
system and ultraviolet background sources result in an average
10\% - 15\% duty cycle~\cite{Scarsi:fy}. Hence, the incredibly
large geometric aperture, $A \approx 7500$~km$^3$ w.e. sr, is
somewhat reduced.

The fluorescence eye consists of several large light collectors (or
telescopes) which image regions of the sky onto clusters of light sensing and
amplification devices. The basic elements of a telescope are the
diaphragm, which defines the telescope aperture, the spherical mirror that must be
dimensioned to collect all the light entering the diaphragm in the
acceptance angular range, and the camera which consists of an array of
photomultiplier tubes (PMTs) positioned approximately on the mirror focal
surface.  The PMTs effectively pixelize the region of the sky covered by the telescope.
The shower development is detected as a long, rather narrow sequence of hit PMTs.
As the point-like image of the shower proceeds through an individual PMT, the
signal rises, levels off, and falls again.

 The sensitivity of the detector depends primarily on the signal ($S$) to
noise ($N$) ratio~\cite{Baltrusaitis:mx},
\begin{equation}
\frac{S}{N} = \frac{N_e N_\gamma}{8\pi} \,\,\left(\frac{c}{\langle B \rangle}\right)^{1/2}\, \,\frac{\kappa_1\,
\,\kappa_2}{R_p^{3/2}}\,\,
\left(\frac{\epsilon D^3}{d}\right)^{1/2}\,\,
\end{equation}
where $\langle B \rangle \approx 400$~m$^{-2}$ sr$^{-1}$
ns$^{-1}$ is the average photon night glow background in the
ultraviolet band~\cite{Catalano}, $\epsilon \approx 20\%$ is the
quantum efficiency of the detector, $\kappa_1 \approx 1$ is the
transmission coefficient in the ozone layer, $\kappa_2 =
e^{-r/\lambda_R}$ is the transmission coefficient in the
atmosphere, $\lambda_{\rm R} \approx 23$~km (at 400~nm) is the
Rayleigh scattering length at sea level, $R_p$ is the distance of
closest approach between the shower and the detector, $r \approx
15$~km is the effective slant depth of the lower atmosphere, $D =
2.5$~m is the diameter of the mirror aperture, and $d = 5$~mm is
the diameter of the PMT. The fluorescence trail is emitted
isotropically with an intensity that is proportional to the
number of charged particles in the shower, $N_e$. The fluorescent
signal is roughly $N_\gamma \sim 4$~photons/electron/m and  $N_e
\approx 0.625\, (E/{\rm GeV}).$ This translates into a $4\sigma$
energy threshold of
\begin{equation}
E = 2.7 \times 10^6 \,\,  R_p^{3/2}\,\,  \exp(r/\lambda_R) \,\,{\rm GeV}\,\,.
\end{equation}
In addition to the fluorescence process, the electrons produce a large photon signal
from \v Cerenkov radiation that is primarily beamed in the forward direction. The production
of fluorescence light is less than $10^{-4}$ of that in the \v Cerenkov cone~\cite{Boley}. For
Earth-penetrating showers emerging upward in the direction of the orbiting telescope, this  \v Cerenkov
signal extends the $4\sigma$ threshold of EUSO to the PeV energy band.

\section{Cross sections}

The cross section for the resonant scattering $\tilde\chi + q(\bar q)\rightarrow
\tilde q (\bar{\tilde q})\rightarrow $ all~\cite{sl}
\begin{eqnarray}
\sigma_{_{\tilde\chi N}} &  = & \sum_{q \bar q} 2\pi \int dx \,\,q(x)\,
\frac{1}{4p_q.p_{\tilde\chi}} \nonumber \\
 & \times & \frac{1}{4} \sum_{\rm spin} |\mathfrak{M}|^2 \delta \left[2(p_q.p_{\tilde\chi}) -
m^2_{\tilde q}\right]\,\,,
\end{eqnarray}
where $p_q$ and $p_{\tilde\chi}$ are the 4-momenta of the quark and neutralino
in the interaction,
\begin{equation}
\frac{1}{4} \sum_{\rm spin} |\mathfrak{M}|^2 = \frac{1}{4}\,|[a_{q_{\rm L}}|^2
+ | a_{q_{\rm R}}|^2]\,m^2_{\tilde q}
\end{equation}
where we have included the contributions from the left-handed and right
handed squarks assuming they are degenerate.
In the above the first sum runs over all quark flavors with parton distribution functions (pdf's) indicated by
$q(x)$. For small L-R mixing, and ignoring quark masses and small
Higgsino mixings one has~\cite{hg}
\begin{eqnarray}
\frac{a_{u_{\rm L}}}{\sqrt{2}} & = & (X_2 \cos \theta_W - X_1 \sin \theta_W) \left(\frac{1}{2} - \frac{2}{3}
\sin^2 \theta_W\right) \nonumber \\
  & \times &\frac{g_2}{\cos \theta_W}+ \frac{2}{3}\, e\, (X_1 \, \cos \theta_W + X_2 \sin \theta_W) \,,
\label{s1}
\end{eqnarray}

\begin{eqnarray}
\frac{a_{u_{\rm R}}}{\sqrt{2}} & = & \frac{2}{3}
\frac{g_2}{\cos \theta_W} \sin^2 \theta_W \,(X_2 \cos \theta_W - X_1 \sin
\theta_W) \nonumber \\
 & + &  \frac{2}{3}\, e\, (X_1 \, \cos \theta_W + X_2 \sin \theta_W)
\label{s2}
\end{eqnarray}
\begin{eqnarray}
\frac{a_{d_{\rm L}}}{\sqrt{2}} & = & (X_2 \cos \theta_W - X_1 \sin \theta_W)
\left(\frac{1}{3} \sin^2 \theta_W -\frac{1}{2}\right) \nonumber\\
 & \times &
\frac{g_2}{\cos \theta_W}
- \frac{1}{3}\, e\, (X_1 \, \cos \theta_W + X_2 \sin \theta_W) \,\,,
\label{s3}
\end{eqnarray}
\begin{eqnarray}
\frac{a_{d_{\rm R}}}{\sqrt{2}} & = &  \frac{1}{3}
\frac{g_2}{\cos \theta_W} \,\sin^2 \theta_W \,(X_2 \cos \theta_W - X_1 \sin \theta_W) \nonumber \\
  & - & \frac{1}{3}\, e\, (X_1 \, \cos \theta_W + X_2 \sin \theta_W)
\label{s4}
\end{eqnarray}
and similar relations hold for the  charm and the strange squarks.
In Eqs.~(\ref{s1}), (\ref{s2}), (\ref{s3}), and (\ref{s4}) $g_2$
is the weak SU(2) gauge coupling constant, $\theta_W$ is the weak
mixing angle, $e=g_2\ \sin\theta_W$ is the U(1)$_{\rm em}$
charge, and $X_1, X_2$ are the projection of $\tilde\chi$ on the
Bino  and Wino super partners, respectively, and we have ignored
small Higgsino components $X_3$ and $X_4$ where $\sum_{i=1}^{4}
|X_i|^2=1$. In the analysis we use renormalization group
evolution of the soft parameters  from the grand unification
scale $M_G$ to low energy to compute the neutralino mass matrix.
This matrix depends on the sign of $\mu$. The recent experiment
on the anomalous magnetic moment of the muon\cite{Bennett:2004pv}
indicates a positive sign for $\mu$ for the supersymmetric
contribution\cite{chatto} and thus in this analysis we have
chosen a positive  $\mu$ sign. The diagonalization of the
neutralino mass matrix determines the projections $X_1, X_2$
which are computed in terms of the input mSUGRA parameters. In
the parameter space of interest, i.e., where $m_{\tilde q}\sim
1$~TeV, and the WMAP constraint is obeyed, the neutralino is
essentially a Bino, and one has $X_1 \simeq 1$ while $X_2,
X_3,X_4$ are relatively small.

 Through use of the parton model relation
$2\,p_q.p_{\tilde\chi}=2\,xP_N.p_{\tilde\chi}$ $(P_N$ is the nucleon
momentum) and the $\delta$-function, one finds the compact
expression
\begin{equation}
\sigma =  \frac{\pi}{4} \sum_{q \bar q} [|a_{q_{\rm L}}|^2 + |a_{q_{\rm R}}|]^2
\,\frac{1}{m^2_{\tilde q}}\, x\,q(x) \,,
\end{equation}
where, using the lab relation $p_N . p_\chi = M_N E_\chi,$ we now
have
\begin{equation}
x= \frac{m^2_{\tilde q}}{2 M_N E_\chi} \,\,.
\end{equation}
We use the CTEQ6D pdf's~\cite{Pumplin:2002vw}, which in the energy region of interest
for up, down, strange and
charmed (quarks and anti quarks)
can be parametrized as
\begin{equation}
x \,q(x) \approx 0.152\,\, x^{-0.382} \,,\,\,\, {\rm with}\,\, 10^{-6}< x < 10^{-4.7} \,.
\end{equation}
Then, for
\begin{equation}
10^{9} \left( \frac{m_{\tilde q}}{1~{\rm TeV}}\right)^2\, <
\frac{E_{\tilde \chi}}{1~{\rm GeV}} < 10^{11.7} \,\,
\left( \frac{m_{\tilde q}}{1~{\rm TeV}} \right)^2\,\,,
\end{equation}
the neutralino-nucleon interaction cross section becomes
\begin{eqnarray}
\sigma_{_{\tilde \chi N}} & = & 1.73 \times 10^{-37} \,
\left( \frac{1~{\rm TeV}}{m_{\tilde q}}\right)^{2.784} \,
\left(\frac{E_{\tilde \chi}}{1~{\rm GeV}}\right)^{0.382} \nonumber \\
 & \times & \sum_{q\bar q} \left[|a_{q_{\rm L}}|^2 + |a_{q_{\rm R}}|^2 \right] \,\, {\rm cm}^2.
\label{crosssection}
\end{eqnarray}

We comment briefly on the relation of our calculation to a
previous estimate~\cite{Berezinsky:1997sb}. In contrast to that
work, we do not characterize the SUSY parameter space according
to the decay branching ratio $\Gamma(\tilde q\rightarrow q \tilde
g)/\Gamma(\tilde q\rightarrow q \tilde\chi).$ The total cross
section, as given above, is independent of any branching to
gluinos. Comparison of Eq.~(\ref{crosssection}) with the cross
section for the competing process $\tilde\chi\ g\rightarrow
\tilde q q$ estimated in~\cite{Berezinsky:1997sb} shows that the
resonant cross section is about an order of magnitude greater.

\section{Event rates}

The most popular mechanism to date to produce ultra-high energy
neutralinos is annihilation or decay of super massive
($10^{12}~{\rm GeV} \alt m_X \alt 10^{16}~{\rm GeV}$)
$X$-particles. To maintain an appreciable decay rate today, one
has to tune the $X$ lifetime to be longer (but not too much
longer) than the age of the universe~\cite{Berezinsky:1997hy,Kuzmin:1997cm}, or
else ``store'' short-lived $X$-particles in topological vestiges
of early universe phase transitions~\cite{Hill:1986mn}. The
cascade decay to cosmic ray particles is driven by the ratio  of
the volume density of the $X$-particle ($n_X = \rho_c
\,\Omega_X\,/m_X$) to its decay time  ($\tau_{_X}$).  This is
very model dependent, as neither the cosmic average mass density
contributed by the relics $(\Omega_X),$ nor $\tau_{_X}$ is  known
with any degree of confidence  ($\rho_c \approx 1.05 \times
10^{-4}\, h^2$ GeV cm$^{-3}$, with $h \equiv$  Hubble constant in
units of 100~km  sec$^{-1}$ Mpc$^{-1}$). Moreover, the internal
mechanisms of the decay and the detailed dynamics of the first
secondaries do depend on the exact nature of the particles.
Consequently, no firm prediction on the expected flux of
neutralinos can be made. However, if there are no new mass scales
between $M_{\rm SUSY} \sim 1$~TeV and $m_X,$ the squark and
sleptons would behave like their corresponding supersymmetric
partners, enabling one to infer from the ``known'' evolution of
quarks and leptons the gross features of the $X$-particle
cascade: the quarks hadronize producing jets of hadrons
containing mainly pions together with a 3\% admixture of
nucleons~\cite{Coriano:2001rt}. The final spectrum is then
dominated by gamma rays and neutrinos produced via pion decay.

In light of the mounting evidence that ultra-high energy cosmic
rays are not gamma rays~\cite{Anchordoqui:2002hs}, proton
dominance at ultra-high energies is achieved by efficient
absorption of the dominant ultra-high energy photon flux on the
universal and/or Galactic radio background. This results in a
recycling of the photon energy down to the MeV -- GeV region.
Thus, since the baryon and photon components of the $X$-particle
decay are correlated by the dynamics, the measurement of the GeV
diffuse gamma ray flux can  significantly constrain the cosmic
ray production by $X$-particles, integrated over
redshift~\cite{Protheroe:1996pd}. In this direction, the new
EGRET limits~\cite{Strong:2003ex} on the photon flux in the GeV
region have severely limited~\cite{Semikoz:2003wv} the
contribution of $X$-particle decay to the  high energy end of the
cosmic ray spectrum reported by the AGASA
Collaboration~\cite{Takeda:2002at}.

\begin{figure}
\begin{center}
\includegraphics[height=8cm]{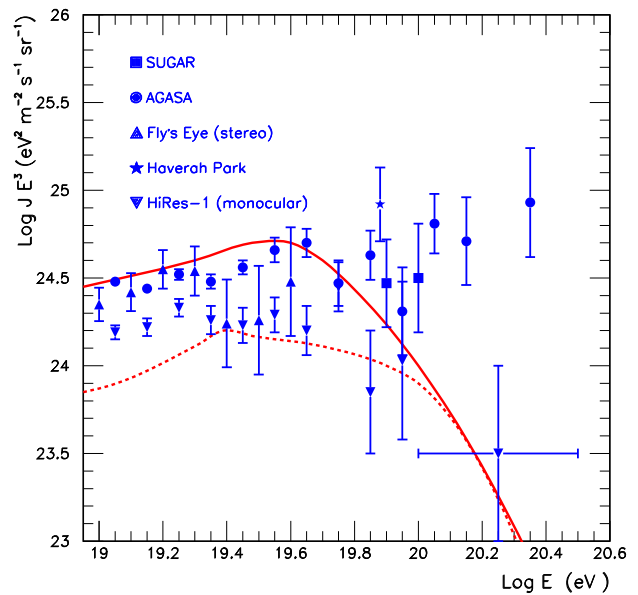}
\caption{The solid line is a 2-component prediction of the ultra
high energy proton flux, consisting of emission from ultra-high
energy ``stars'' plus decay of super heavy relics, both sources
distributed uniformly throughout the
universe~\cite{Barbot:2002kh}. The dashed line indicates the
contribution from the $X$-decay, with initial state of 5 quarks +
5 squarks and $m_X = 2 \times 10^{12}$~GeV. The upper end of the
spectrum as seen by different experiments
(AGASA~\cite{Takeda:2002at}, HiRes~\cite{Abu-Zayyad:2002sf},
Fly's Eye~\cite{Bird:wp} SUGAR~\cite{Anchordoqui:2003gm}, and
Haverah Park~\cite{Ave:2001hq}) is also shown for comparison.}
\label{fig1}
\end{center}
\end{figure}

\begin{figure}
\begin{center}
\includegraphics[height=8cm]{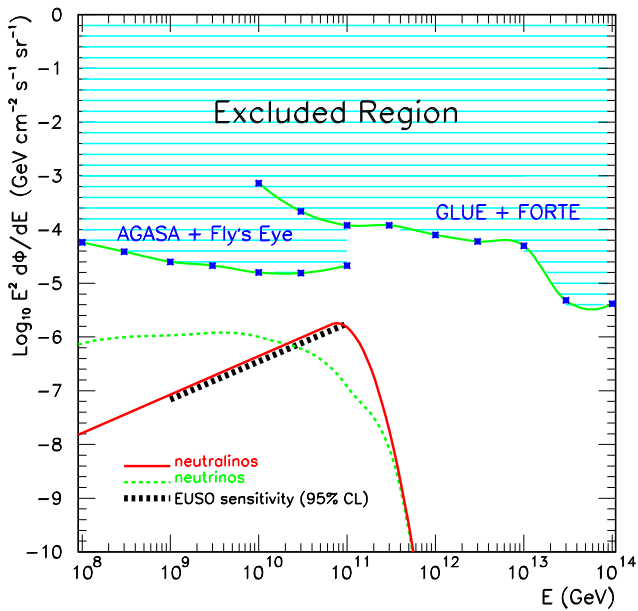}
\caption{The solid (dashed) line indicates the
neutralino~\cite{Barbot:2002et} (neutrino plus
anti-neutrino~\cite{Barbot:2002kh}) energy spectrum corresponding
to the proton flux from $X$-particle decay given in
Fig.~\ref{fig1}. The horizontally-hatched region at the top of the
figure has been already excluded from null results at AGASA +
Fly's Eye (95\% CL)~\cite{Anchordoqui:2002vb} and GLUE + FORTE
(90\% CL)~\cite{Gorham:2001aj}. The thick dashed line indicates
the 95\% CL sensitivity of the EUSO mission.} \label{fig2}
\end{center}
\end{figure}

In our analysis we adopt the recent estimates of neutralino fluxes
from decay of super heavy relics derived in
Ref.~\cite{Barbot:2002et}. The normalization is determined by
matching the $X$-particle baryon flux to the difference between
the observed spectrum at $E \approx 10^{11}$~GeV and
contributions from a homogeneous population of astrophysical
sources. Among the homogeneous models discussed in
Ref.~\cite{Barbot:2002et}, the only one to accommodate both the
ultra-high energy cosmic ray intensity and the GeV photon flux  is
a  distribution of $X$ particles with $m_X\approx 2 \times
10^{12}$~GeV and primary 10-body decay $X \rightarrow 5q\ 5\tilde
q.$ As can be seen in Fig.~\ref{fig1}, this model is constrained
to fit only the AGASA data below $10^{11}$~GeV, with the baryonic
flux from $X$-decay contributing less than 1/3 of the
total~\cite{Barbot:2002kh}. It is noteworthy that the associated
flux of neutrinos from $\pi^\pm$ decay produced in this scenario,
is also consistent with existing data. This is shown in
Fig.~\ref{fig2}.

As mentioned in the Introduction, in order to comply with
existing fits of mSUGRA to the WMAP dark matter density, the
contribution to this density from superheavy relics must be much
less than that of the relic neutralinos. We recall that the
cosmic ray flux resulting from $X$-particle decays depends on
the dimensionless parameter $r_X\equiv \xi_Xt_0/\tau_{_X}$, where
$\xi_X = \Omega_X/\Omega_{\rm CDM},$ and
$t_0$ is the age of the universe. Scenarios which include
in the rate normalization photon flux  from $X$-particles clustered
in the halo lead to a value $r_X
\sim 5\times 10^{-11}$~\cite{Berezinsky:1997hy}. Omission of the photon channel
in the normalization increases $r_X$ by a factor of about 10, and the
extension from halo dominance to the homogeneous population used in this
work further increases $r_X$ by a factor of 15. Since  models
of $X$-production and decay typically lead to exponential dependence
of both $\xi_X$ and $\tau_{_X}$ on a reheating temperature $T_R$ and a quantum
mechanical tunneling action, respectively, there is no impediment on
accommodating this change in $r_X$ while maintaining
$\Omega_X\ll\Omega_{\rm CDM}.$ (For example, for the model provided
in~\cite{Berezinsky:1997hy}, $\xi_X \propto e^{-2m_X/T_R}\sim 10^{-4} - 10^{-8}.$)

Establishing a  neutralino signal in upward going showers will
require a suppression of neutrino events which create a
background. This occurs naturally because the difference in
neutrino and neutralino  cross sections leads to differing
absorption lengths in the Earth's crust. The neutrino flux is
greatly attenuated by selecting events entering the Earth at
angles $>5^{\circ}$ below the horizon. For the EUSO effective
aperture (duty cycle of 10\%), the neutrino background from a
homogeneous population of $X$'s decaying into  $5q\ 5\tilde q$
(during the 3-yr mission lifetime) is about 0.3 events~\cite{bcd}.
The use of  Poisson statistics implies the observation of  $\ge 3.09$
events establishes a signal significance at 95\%
C.L.~\cite{Feldman:1997qc}.

\begin{figure}
\begin{center}
\includegraphics[height=8cm]{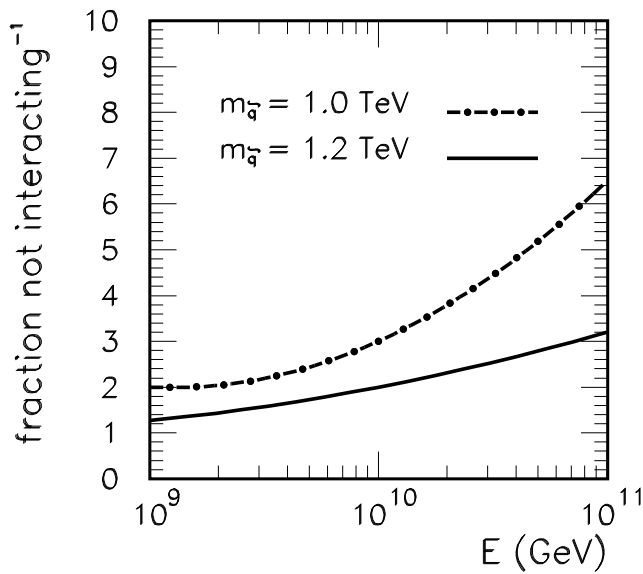}
\caption{Fraction of neutralinos which traverse the Earth, for
all zenith angles less than 85$^\circ$, as a function of energy.}
 \label{fig3}
\end{center}
\end{figure}

The determination of a neutralino event rate will depend strongly
on the neutralino-nucleon cross section, and hence, through
Eq.~(\ref{crosssection}), on the squark mass. Let us define $P$
as the probability that the neutralino does not interact while
traversing the Earth's crust.  A large cross section will reduce
$P$ and lead to a lowering of emerging flux. In Fig.~\ref{fig3} we
illustrate the behavior of $P$ with energy, for two different
values of $m_{\tilde q}$.  On the other hand, it will enhance the
event rate in the atmosphere for the neutralinos which do emerge
without interacting. (Conservatively, we omit regeneration
effects in the Earth when estimating a signal.) Apart from the
direct effect on the size of the cross section, the squark mass
will also influence the event rate through the position of the
resonant peak: a larger squark mass will probe higher value of
$E_{\tilde\chi}$, and hence lead to a decrease in event rate
because of the sharply falling flux. After considering these
effects, we have found that a value of $m_{\tilde q}\simeq 1$~TeV
leads to an optimal signal. The degree of tolerance is fairly
narrow: values  lower than 800 GeV or larger than 1200 GeV will
vitiate a signal at 95\% C.L. We note once more that this
coincides with the region of Bino-dominance for the parameter
space compatible with WMAP data~\cite{wmap1}.

For a given  flux of neutralinos
$d\Phi/dE_{\tilde \chi}$ and observation time $t$, the event rate at EUSO is
found to be
\begin{equation}
{\cal N} = \int_{E_{\tilde \chi}^{\rm min}}^{E_{\tilde \chi}^{\rm max}}
dE_{\tilde \chi}\,\,N_A\,\,P\,\, \frac{d\Phi}{dE_{\tilde \chi}}\,
\sigma_{_{\tilde \chi N}}\, A\,\,\epsilon_{_{\rm DC}}\,\,t,
\end{equation}
where $N_A$ is Avogadro's number and $\epsilon_{_{\rm DC}} \approx
10\%$ is the duty cycle. The fraction of unscathed neutralinos,
integrated over zenith angle $< 85^\circ$, emerging upward
in the direction of the orbiting telescope is
found to be $P \approx 107\,\, (E_{\tilde\chi}/{\rm GeV})^{-0.256}$. Now, by
setting $E_{\tilde \chi}^{\rm
min} = 10^{9}$~GeV and $E_{\tilde \chi}^{\rm max} = 10^{11}$~GeV,
one finds that for the full mission lifetime EUSO will typically
collect between 4 and 5 events, and hence provide sensitivity to
the neutralino flux given in Fig.~\ref{fig2} at more than 95\% CL.

\section{Conclusions}

Using a technique that exploits the different absorption lengths
 of neutrinos and neutralinos in the Earth's crust, we have
estimated the sensitivity of EUSO to isolate upward coming
showers of ultra-high energy $\tilde\chi$. The neutralino-nucleon
interaction  has been approximated by resonant squark production,
with the neutralino being largely Bino in composition, and
$m_{\tilde q}\simeq$ 1 TeV. We have shown that, during the
complete mission lifetime, the telescope will be sensitive to
$E_{\tilde \chi}^2\ d\Phi/dE_{\tilde \chi} > 1.1 \times 10^{-6}\
(E_{\tilde \chi}/{\rm GeV})^{0.7}$ GeV cm$^{-2}$\ yr$^{-1}$
sr$^{-1}$ at the 95\% CL, for $10^{9}~{\rm GeV} < E_{\tilde \chi}
< 10^{11}~{\rm GeV},$ and for the region $m_{\tilde q}=1.0\pm
0.2$ TeV. A hard spectrum $\propto E_{\tilde\chi}^{-1.3}$ is
typical of super heavy relic $N$-body decays that are purely
hadronic. This is a conservative estimate, since regeneration
effects have been only considered in computing the neutrino
background. We have explicitly analyzed the case in which the
flux of ultra-high energy neutralinos is produced via decay of
metastable heavy ($m_X = 2 \times 10^{12}$~GeV) particles with
uniform distribution throughout the universe, and primary decay
mode into 5 quarks + 5 squarks. The normalization of
$n_X/\tau_X$  has been fixed to contribute about 1/3 of the
events reported by the AGASA Collaboration below
$10^{11}$~GeV~\cite{Takeda:2002at}. For this particular case,
EUSO will collect between 4 and 5 neutralino events (with 0.3 of
background) in $\approx 3$~yr of running.

Existing limits on the diffuse photon flux in the GeV region
strongly limit the sensitivity of EUSO for primary 2-body decays
of hadronic nature. This is because, for the normalization of the
baryonic contribution to the ultra-high energy cosmic ray flux
assumed in Ref.~\cite{Barbot:2002et}, which is marginally
consistent with new EGRET bounds, the accompanying neutralino flux
produced for $q \bar q$ and $q \tilde q$  is about an order of
magnitude below the flux generated in the primary 10-body decay
discussed above. On the other hand, the telescope can still be
sensitive to the leptonic mode $X \rightarrow l \tilde l.$ In
this case, by reducing $n_X/\tau_X$ by a factor of $\sim$ 2, one
lessens the problem with EGRET data and still leaves a window
open for neutralino detection at EUSO.

We note that for a Bino-like neutralino, the primary decay mode
(90\% branching fraction) of the squark is $\tilde q\rightarrow q
\tilde g,$ with a subsequent decay $\tilde g\rightarrow q\bar q \
\tilde\chi.$ Thus, the neutralino energy of the decay is about
1/6 of the primary energy. In the remaining 10\% of the decays,
$\tilde q\rightarrow q \tilde\chi.$ In either case, the shower
energies are far above the $\sim$ 1 PeV threshold for the
detector. If more detailed considerations are warranted in the
future, regeneration effects during passage through Earth can be
assessed, taking into account the energy losses of the decay
modes. These effects will lead to some enhancement of $P$, and
consequently of the event rate.

We turn now to a brief discussion on the potential of the planned NASA
mission Orbiting Wide-angle Light-collectors
(OWL)~\cite{Stecker:2000ek}. This mission will involve
photo detectors mounted on 2 satellites in low equatorial orbit
(600 - 1200 km). The eyes of the OWL will stereoscopically image
a geometric area of $\sim 9 \times 10^5$~km$^{2},$ yielding
$A \sim 3 \times 10^{6}$~km$^2$ sr. With its
superior effective aperture ($\epsilon_{_{\rm DC}} \approx
10\%$), a 10 yr mission lifetime will allow one to discern on
contributions of metastable relics to the upper end of the cosmic
ray spectrum at the level of 1 part in $10^2.$  Consequently, the
data from this mission will allow one to probe more deeply the parameter space
of mSUGRA and its extensions.

\acknowledgments{We would like to thank Michael Kachelriess, Dmitri Semikoz,
and Guenter Sigl for valuable discussions. This work has been partially
supported  by the U.S. National Science Foundation (NSF); grants
No.\ PHY--0140407  (LA), No.\ PHY--0244507 (HG), and No.\ PHY--0139967 (PN).


\begin{thebibliography}{999}
\bibitem{sl} There will be also resonant selectron
scattering of neutralinos from electrons. This process
probes a narrow region of the neutralino flux spectrum,
and contributes negligibly to the event rate.
\end{thebibliography}
\end{document}